\documentclass[fleqn,usenatbib]{mnras}
\usepackage{newtxtext,newtxmath}
\usepackage[T1]{fontenc}
\usepackage{fix-cm}
\DeclareRobustCommand{\VAN}[3]{#2}
\let\VANthebibliography\thebibliography
\def\thebibliography{\DeclareRobustCommand{\VAN}[3]{##3}\VANthebibliography}

\usepackage{graphicx}	% Including figure files
\usepackage{amsmath}	% Advanced maths commands

\title[The efficient star-forming regions of SESNe]{The efficient star-forming regions of stripped-envelope supernovae}

\author[M. Solar et al.]{
Martín Solar,$^{1}$\thanks{E-mail: martin.solar@amu.edu.pl}
Michał~J. Michałowski,$^{1}$\thanks{E-mail: michal.michalowski@amu.edu.pl}
Jakub Nadolny,$^{1,2,3}$
Lluís Galbany,$^{4,5}$
Joseph~P. Anderson,$^{6}$\newauthor
Jesper Sollerman,$^{7}$
Thallis Pessi,$^{6}$
Emmanouil Zapartas,$^{8,9}$
Bruno \v{S}laus,$^{1}$
Jason Alexander,$^{1}$\newauthor
Oleh Ryzhov,$^{1}$
Jens Hjorth,$^{10}$
Cristina Jiménez-Palau,$^{4,5}$
Przemysław Nowaczyk,$^{1}$\newauthor
Devang Somawanshi,$^{1}$
and Aleksandra Leśniewska$^{10,1}$
\\
$^{1}$Astronomical Observatory Institute, Faculty of Physics and Astronomy, Adam Mickiewicz University, ul. Słoneczna 36, 60-286, Poznań, Poland\\
$^{2}$Instituto de Astrofísica de Canarias, E-38205 La Laguna, Tenerife, Spain\\
$^{3}$Departamento de Astrofísica, Universidad de La Laguna (ULL), E-38205 La Laguna, Tenerife, Spain\\
$^{4}$Institute of Space Sciences (ICE, CSIC), Campus UAB, Barcelona, Spain\\
$^{5}$Institut d’Estudis Espacials de Catalunya (IEEC), Barcelona, Spain\\
$^{6}$European Southern Observatory, Alonso de Córdova 3107, Vitacura, Casilla 19001, Santiago, Chile\\
$^{7}$The Oskar Klein Centre, Department of Astronomy, Stockholm University, Albanova University Center, Stockholm, Sweden\\
$^{8}$Physics Department, National and Kapodistrian University of Athens, 15784 Athens, Greece\\
$^{9}$Institute of Astrophysics, Foundation for Research and Technology-Hellas, 71110 Heraklion, Greece\\
$^{10}$DARK, Niels Bohr Institute, University of Copenhagen, Jagtvej 155A, DK-2200 Copenhagen N, Denmark\\
}

\date{Accepted XXX. Received YYY; in original form ZZZ}

\pubyear{\the\year{}}

\begin{document}
\label{firstpage}
\pagerange{\pageref{firstpage}--\pageref{lastpage}}
\maketitle

\begin{abstract}
Massive stars ($\geq8\,\rm{M}_{\odot}$) play a key role in shaping the interstellar medium of galaxies through stellar feedback.
However, how these stars form and evolve before exploding as core-collapse supernovae (SNe) remains elusive.
We compute for the first time the star-formation efficiencies (SFEs) at the locations of hydrogen-rich (H-rich) SNe and stripped-envelope SNe (SESNe) to constrain their progenitor properties.
We used VLT/MUSE and ALMA observations of $\mathrm{H}~\alpha/\mathrm{H}~\beta$ and CO(2--1) emission lines to trace the components of the warm ionised gas and cold molecular gas, respectively.
% Both observations resolve individual \ion{H}{ii} regions and giant molecular clouds at spatial resolutions on cloud-scales ($\sim 100$~pc; at distances below 80~Mpc).
Both observations resolve individual \ion{H}{ii} regions and giant molecular clouds at spatial resolutions on cloud-scales ($\sim 100$~pc).
This combined data allows us to compute the SFE from the star formation rate (SFR) and the molecular gas mass ($M_{\rm mol}$) as $\mathrm{SFE}=\mathrm{SFR}/M_{\rm mol}$.
We find that SESNe explode in environments that are currently forming stars eight times more efficiently than those of H-rich SNe (higher SFR for SESNe with similar $M_{\rm mol}$).
% {\color{cyan}This is consistent with the scenarios in which the majority of SESNe are produced from very massive stars ($\geq 20~\rm{M}_{\odot}$) and in which the majority are produced from interacting binaries ($\leq 20\,\rm{M}_{\odot}$).
On one hand, this is consistent with the scenario in which the majority of SESNe are produced from very massive stars ($\geq20~\rm{M}_{\odot}$) if the initial mass function is top-heavy.
On the other hand, most of SESN progenitor channels are formed from interacting binaries ($\leq20~\rm{M}_{\odot}$) if an increased binary system formation rate is connected with turbulences and, in turn, with the boost to SFE.
Then, an increased binary fraction could explain the enhanced H~$\alpha$ luminosities.
% We do not rule out a bimodal distribution of initial masses for SESN progenitors.
% On one hand, high SFR is correlated with the production of very massive stars and/or interacting binary systems.
% On the other hand, for a fixed amount of $M_{\rm mol}$, the binary fraction could be enhanced at higher SFRs.}
% We do not rule out a bimodal distribution of initial masses for SESN progenitors.
In summary, SESNe preferentially occur in regions of intense, efficient star formation rather than simply higher gas content. 
\end{abstract}


%%%%%%%%%%%%%%%%%%%%%%%%%%%%%%%%%%%%%%%%%%%%%%%%%%

%%%%%%%%%%%%%%%%% BODY OF PAPER %%%%%%%%%%%%%%%%%%

%---INTRODUCTION---
\section{Introduction}\label{sec:sec1}

Core-collapse supernovae (CCSNe) are explosions of stars within initial masses higher than $8~\rm{M}_{\odot}$ (hereafter massive stars).
Understanding these explosions is important because they shape the interstellar medium (ISM) through different mechanisms. 
For instance, CCSNe synthesise metals and dust, inject momentum causing both positive and negative feedback on star formation, generate cosmic rays at evolved stages, and produce neutrinos \citep{2017Branch}. 
As massive stars are not static objects, they have a complex formation and evolutionary phase, leading to different types of CCSNe \citep{2019Modjaz,2025Gilkis,2026Souropanis}.
Therefore, it is necessary to constrain the properties of CCSN progenitors to predict how they explode.

Historically, supernovae (SNe) were classified as Type I and Type II based on their spectral features at the time of maximum optical light \citep{1941Minkowski}.
In particular, CCSNe are classified as follows: Type II SNe are hydrogen-rich (H-rich), whereas stripped-envelope SNe (SESNe) lack hydrogen and silicon features.
SESNe are further subdivided into several subclasses:
Type Ib SNe show no hydrogen lines; Type IIb SNe initially resemble Type II SNe shortly after explosion, but after several weeks their spectra evolve to resemble those of Type Ib; and Type Ic SNe exhibit neither hydrogen nor helium lines \citep{1997Filippenko,2017GalYam}.
Ideally, the classification of CCSNe should be linked to the nature of their progenitor stars, however, the observational classification alone does not provide a complete physical interpretation of their origin.

One powerful technique that enables one to constrain SN progenitors is the statistical study of their environments \citep[][and references therein]{2015Anderson}.
This approach has been used across ultraviolet \citep[UV;][]{2006Fruchter,2013Kangas,2023bSun}, optical \citep{2011Leloudas,2013Kuncarayakti,2013Lyman,2017Kangas,2018Kuncarayakti,2014Hakobyan,2016Galbany,2023Pessi,2024Thone,2025Ganss}, submillimetre \citep{2017Galbany}, and radio \citep{2015Michalowski,2018bMichalowski,2020Michalowski,2020bMichalowski,2015Arabsalmani,2022Arabsalmani,2024deUgartePostigo,2026Slaus} observations.
However, these previous studies were limited to small samples, observations at galactic-scale spatial resolutions ($> 1~\mathrm{kpc}$) or integrated observations of entire galaxies.
To constrain the SN progenitor channels, it is essential to resolve the stellar populations or gas properties at the explosion sites.
Although SNe themselves and galaxy-averaged host properties have been studied extensively \citep{2008Hakobyan,2009Boissier,2010Arcavi,2012Kelly,2024Jones,2024Qin,2024Pritchet,2026Nugent}, higher-resolution observations are necessary to understand the nature of SNe.
Cloud-scale observations \citep[$\sim$\,$100~$pc;][]{2016Leroy} do not resolve individually star-forming regions \citep[\ion{H}{ii};][]{2003Oey} or giant molecular clouds \citep[GMCs;][]{2017MivilleDeschenes}, but these spatial scales recover their general trends \citep[][and references therein]{2024Schinnerer}.
The resolution comparable to typical sizes of \ion{H}{ii} regions and GMCs is beneficial as CCSN progenitors should be associated with the ionised gas and their parent molecular clouds at the moment of the explosions.

The advent of integral field unit \citep[IFU;][]{2012Sanchez,2015Bundy} spectrographs revolutionised studies of SN environments.
Specifically, the Multi Unit Spectroscopic Explorer (MUSE) at the Very Large Telescope \citep[VLT; hereafter VLT/MUSE;][]{2010Bacon} is able to physically resolve the warm ionised gas (typically $10^4$~K) conditions (e.g., star formation rate, metallicity, dust extinction, etc.) at explosion sites and infer the ISM properties in which progenitor stars were born.
For example, the optical range includes Balmer series emission lines, making it possible to trace the warm ionised gas.
Since the most massive stars are responsible for ionising the surrounding gas, the typical lifetimes of \ion{H}{ii} regions are comparable to the explosion delay time of these massive stars \citep[$\sim$\,10~Myr;][]{2010Goddard,2021Torniamenti,2022Chevance,2026Wainer}.
From optical environmental studies using VLT/MUSE, SESN locations have been associated with regions of higher star formation rates (SFRs) than those of H-rich SNe \citep{2018Kuncarayakti,2023Pessi,2024MaykerChen}.
Thus, SESNe may be associated with younger and more massive star-forming regions, implying progenitors more massive than those of H-rich SNe.
However, all these studies solely used Balmer lines as a proxy to derive the properties of SN progenitors.
Balmer lines only trace recent star formation, and a more promising approach is to include multiwavelength observations of the ISM.

\cite{2017Galbany} served as a benchmark study of cold molecular hydrogen gas ($\sim$10--30~K; hereafter molecular gas) mass ($M_{\mathrm{mol}}$) reservoirs (at $\sim$\,$1.5$~kpc spatial scales) at the SN explosion sites.
Later, using Atacama Large Millimeter/submillimeter Array (ALMA) data at cloud-scale spatial resolutions, \cite{2023MaykerChen} explored molecular gas at SN locations and found modest differences between H-rich SNe and SESNe.
Following a similar approach, \cite{2024Solar} used a larger CCSN sample and found that H-rich SNe and SESNe share similar molecular-gas environmental properties.
Combining both approaches (VLT/MUSE + ALMA) makes it possible to compute the star formation efficiency (SFE) from the ratio of the star formation rate to the amount of molecular gas ($\mathrm{SFE} \equiv \mathrm{SFR}/M_{\rm mol}$, or the molecular gas depletion time $\tau_{\rm{dep}}\equiv\mathrm{SFE}^{-1}$).
This estimator indicates how effectively a GMC converts its molecular gas into stars.

Previous works have already analysed star formation rate and molecular gas properties of host galaxies of various transients.
Examples of such transients include long-duration ($>2$~s) Gamma-Ray Bursts (GRBs; \citealt{2014Hatsukade,2020aHatsukade,2015aStanway,2015bStanway,2016Michalowski,2018aMichalowski,2019Hashimoto,2020Arabsalmani,2020deUgartePostigo,2024deUgartePostigo,2021Chen}; \citealt{2023Nadolny,2024Thone}), short-duration ($<2$~s) GRBs \citep{2021NicuesaGuelbenzu}, and super-luminous SNe \citep{2019Arabsalmani,2020bHatsukade}.
For CCSNe, detailed host-galaxy features have been studied in small samples
\citep[three H-rich SNe and two Ic-BL SNe;][]{2018bMichalowski,2020Michalowski}.
However, there has been a lack of statistical studies on SFE measurements.
In this Paper, we measure for the first time the SFE at the locations of recent CCSN explosions ($< 100$~yr) using a statistical sample, resolved at cloud scales.
For this purpose, we combine VLT/MUSE and ALMA data to measure warm ionised gas and molecular gas, respectively.
We use a flat $\Lambda$-CDM cosmological model with $H_{0} = 69$~km~s$^{-1}$~Mpc$^{-1}$, $\Omega_{\Lambda} = 0.71$, and $\Omega_{m} = 0.29$ \citep{2015Schneider}.

%---METHOD---
\section{Method}\label{sec:sec2}

\subsection{Sample Selection}\label{sec:subsec1}

Only CCSNe are included in the sample; thermonuclear SNe (or Type Ia SNe) were removed.
This is because thermonuclear SNe have different explosion mechanisms and progenitor systems, with longer delay times ($> 100$~Myr), which prevents direct comparison with CCSN progenitor channels at the explosion sites \citep{2015bAnderson}.
The CCSN sample was collected from different databases \citep{1989Barbon, 1993Tsvetkov, 1998Rutledge, 1999Barbon, 2004Tsetkov, 2012Lennarz, 2017Guillochon}. 
The CCSNe were divided into the following classification groups: Type II, IIP, and IIn as H-rich SNe and Type Ib, IIb, Ib/c, Ic, Ic-pec, and Ic-BL as SESNe.

To constrain CCSN progenitors through the SFE of their environments, both measurements of star formation and molecular gas are required.
The spectral lines of the Balmer series for hydrogen-alpha ($\mathrm{H}~\alpha$; at $6562.79~\mathrm{\mathring{A}}$) and hydrogen-beta ($\mathrm{H}~\beta$; at $4861.35~\mathrm{\mathring{A}}$), and carbon monoxide (CO) using the $J = 2\rightarrow1$ rotational transition (hereafter CO[2--1], at $230.54~\mathrm{GHz}$) trace the properties of the star-forming regions and molecular gas, respectively.
We selected available observations from both VLT/MUSE and ALMA at spatial resolutions of $\sim$\,$100$~pc for host galaxies of known CCSNe in the southern hemisphere.
The instrumental limitations constrained the distances of the sources at $<80$~Mpc (redshift $z<0.018$).
Also, to avoid strong projection effects, we used face-on galaxies with low inclination angles \citep[$i<75\degr$;][]{2014Makarov,2020Lang}.
% Also, to avoid strong projection effects, we used face-on galaxies with low inclination angles ($i<75\degr$)
% Inclination angles from PHANGS galaxies were taken from \cite{2020Lang}, and those of the remaining galaxies were taken from the HyperLEDA\footnote{http://atlas.obs-hp.fr/hyperleda/} database \citep{2014Makarov}.

In this Paper, we made use of the following dataset (see Section~\ref{sec:subsec2} for data description):
the Physics at High Angular resolution in Nearby GalaxieS\footnote{https://sites.google.com/view/phangs/home} (PHANGS) survey \citep{2021Leroy_a,2021Leroy_b,2022Emsellem}, the All-weather MUse Supernova Integral-field Nearby Galaxies (AMUSING) survey \citep{2016Galbany,2020LopezCoba}, the ALMA CO SN (ACOS) survey \citep[2021.1.00099.S: PI M. Michałowski;][]{2024Solar}, our own VLT/MUSE observations (114.27G5.001: PI M. Solar), and ALMA archival data\footnote{https://almascience.eso.org/aq/}.

The PHANGS sample comprises 90 nearby galaxies ($<20$~Mpc), of which all were observed by ALMA and 19 by both ALMA and VLT/MUSE.
In these 19 PHANGS-ALMA-VLT/MUSE galaxies, 25 CCSN explosions\footnote{SN~1926A, SN~1961I, SN~1964F, SN~1967H, SN~1972Q, SN~1973R, SN~1979C, SN~1983V, SN~1985P, SN~1986I, SN~1995V, SN~1997bs, SN~1999gn, SN~2001du, SN~2006ov, SN~2009hd, ASASSN-14ha, SN~2014L, SN~2016cok, SN~2017gax, SN~2019ehk, SN~2020jfo, SN~2020oi, SN~2022aau, and SN~2022acko} have been discovered in 10 host galaxies\footnote{NGC~1087, NGC~1300, NGC~1365, NGC~1433, NGC~1566, NGC~1672, NGC~3627, NGC~4254, NGC~4303, and NGC~4321}.
In addition, the AMUSING survey collects VLT/MUSE host galaxy observations at the location of around 1000 SNe.
Data from AMUSING and ALMA archive result in eight CCSNe\footnote{SN~2003jg, SN~2004gt, SN~2004ip, SN~2005at, SN~2007C, SN~2013dk, SN~2016adj, SN~2023dpj} exploding in seven host galaxies\footnote{NGC~2997, NGC~4038, NGC~4981, NGC~5128, NGC~5135, NGC~6744, and PGC~084885}.
% Finally, the cross-match of ACOS and our own VLT/MUSE observations resulted in eight CCSNe\footnote{SN~1994ai, SN~1997dq, SN~2005lr, SN~2011hp, SN~2011jl, SN~2012ap, LSQ13doo, and SN~2014ad} from eight host galaxies\footnote{ESO~440-011, ESO~492-002, NGC~0908, NGC~1729, NGC~3354, NGC~3810, NGC~4219, and PGC~037625} and one source had previous AMUSING data (SN~2009bb in NGC~3278; \citealt{2018bMichalowski}).
Finally, we used supernovae from the ACOS survey and followed up with our own VLT/MUSE observations.
We selected follow-up observations based on the data quality of the ACOS survey (i.e., it is possible to distinguish background noise from the physical signal).
This resulted in eight CCSNe\footnote{SN~1994ai, SN~1997dq, SN~2005lr, SN~2011hp, SN~2011jl, SN~2012ap, LSQ13doo, and SN~2014ad} from eight host galaxies\footnote{ESO~440-011, ESO~492-002, NGC~0908, NGC~1729, NGC~3354, NGC~3810, NGC~4219, and PGC~037625}.
One ACOS source had previous AMUSING data (SN~2009bb in NGC~3278; \citealt{2018bMichalowski}).
The sample consists of 42 SNe in 26 host galaxies.
The number of H-rich SNe and SESNe resulted in 22 and 20, respectively.

% {\color{orange}The PHANGS sample comprises 90 nearby galaxies ($<20$\,Mpc), of which all were observed in ALMA and 19 in both ALMA and MUSE.
% From these 19 PHANGS-ALMA-MUSE galaxies, 25 CCSN explosions have been discovered in 10 host galaxies.
% In addition, the AMUSING survey collects VLT/MUSE host galaxy observations at the location of around 1000 SNe.
% Data from AMUSING and ALMA archive result in eight CCSNe exploding in seven host galaxies.
% Finally, the cross-match of ACOS and our own MUSE observations were eight CCSNe from eight host galaxies and one source had previous AMUSING data.
% \color{orange}The CCSN sample resulted in a total of 42 sources in 26 host galaxies.
% The number of H-rich SNe and SESNe used in this work is 22 and 20, respectively.}

% % The PHANGS galaxies used here have typical stellar masses of $9.80<\log_{10}\,M_{*}\,[\mathrm{M}_{\odot}]<11$ and star formation rates of $-0.02<\log_{10}\,\mathrm{SFR}\,[\mathrm{M_{\odot}/yr}]<0.90$.

\subsection{Data Description}\label{sec:subsec2}

The PHANGS is a survey aimed at studying the cloud-scale structure of spiral, star-forming, main-sequence galaxies using multiwavelength observations.
PHANGS observations covers full galaxy extension in H~$\alpha$, H~$\beta$, and CO(2--1).
% PHANGS galaxies have the advantage that their entire disks are covered.
% {\color{cyan}
% In particular, the PHANGS collaboration offers emission line flux maps of H~$\alpha$, H~$\beta$, and CO(2--1).}
We refer to \cite{2021Pessa} for a description of how the maps were computed.
For observations not covered by PHANGS, we followed a similar approach for archival data in order to use as homogenous datasets as possible.
This subsection describes the archival data, first for VLT/MUSE and later for ALMA.

MUSE \citep{2010Bacon} is a second generation instrument located at the Nasmyth focus of UT4 at the VLT at the Paranal Observatory of the European Southern Observatory (ESO).
The VLT/MUSE field of view comprises 24 IFU modules.
Each of these IFU channels contains 48 slices that disperse the light to create a 3D data cube. 
Specifically, we used the Wide Field Mode with natural seeing, which covers a field of view of $1 \farcm 0 \times 1 \farcm 0$, a spatial sampling of $0\farcs2 \times 0\farcs2$, a spatial resolution of $0\farcs4$ at 700~nm, a wavelength range 465--930~nm, a mean resolution of 3000, and a resolving power of 2000 at 460~nm and 4000 at 930~nm.
VLT/MUSE spectral data was collected from the AMUSING survey (from the specific program 095.D-0172: PI H. Kuncarayakti) and our own observations, and the execution time per target ranged from 1 to 2~h.
Observations from 095.D-0172 were performed during 2015 May.
Observations from 114.27G5.001 were performed from 2024 November to 2025 March (aimed to follow up the ACOS observations, described in the following paragraph).
The VLT/MUSE data was reduced using the ESO Reflex MUSE pipeline \citep{2020Weilbacher}.
All data are publicly available.

ALMA observes at the submillimetre and millimetre wavelengths.
It is composed of 66 antennas (54 of 12-meter diameter and 12 of 7-meter diameter).
For ALMA, we made use of data that covered the CO(2--1) line at resolutions of $100$~pc, observable at its Band 6 \citep[$\sim$\,$211$--$275$~GHz, $\sim$\,$1.1$--$1.4$~mm;][]{2004Ediss,2009Wootten}.
For the case of ACOS, the sensitivity was set to a  root mean square of 100~mK and a bandwidth of 2.049~km~s$^{-1}$ (1.6~MHz) per 50~pc beam.
The observations are centred at the SESN position to gain the best signal-to-noise ratio.
The observed integration time ranged from 0.3 to 3.6~h, and requested angular resolutions of $0\farcs5$ to $1\farcs1$.
The sensitivity to the large-scale structures (on kiloparsec scales) is not required for the detection of marginally-resolved sources.
The dual polarisation mode was used to obtain the best possible sensitivity.
Observations from 2021.1.00099.S were made from 2022 January to 2022 May.
% For ALMA archival data, we selected observations with spatial resolutions able to resolve the GMC scales.
For ALMA archival data\footnote{2013.1.00243.S, 2015.1.00902.S, 2015.1.00956.S, 2017.1.00886.L, 2018.A.00062.S, 2018.1.00272.S, and 2019.1.01305.S}, angular resolutions ranged from $0\farcs22$ to $5\farcs77$, able to spatially resolve cloud-scale observations.
The ALMA data was reduced and analysed using the Common Astronomy Software Applications \citep{2022CASATeam}.
All data are publicly available.

\subsection{Star Formation Efficiency at Core-collapse Supernova Explosion Sites}\label{sec:subsec3}

The main goal of this Paper is to analyse the local ISM conditions through the SFE for CCSNe.
This parameter is calculated from the ratio of the surface densities of SFR ($\Sigma_{\rm{SFR}}$) and M$_{\mathrm{mol}}$ ($\Sigma_{\rm{mol}}$).
VLT/MUSE data was corrected for foreground Galactic extinction using the NASA/IPAC Infrared Science Archive\footnote{https://irsa.ipac.caltech.edu/applications/DUST/} \citep{2011Schlafly}.
To calculate the SFR from H~$\alpha$ and H~$\beta$, we first subtracted the optical continuum.
For each spectrum, we fitted the best stellar continuum with \texttt{STARLIGHT} \citep{2005CidFernandes} and subtracted it from the observed spectrum to obtain a pure gas-phase emission spectrum.
The integrated area under the Gaussian profile fitted to the respective emission lines gave us an estimate of the flux.

The SFR measured from hydrogen recombination cascades is sensitive to the effects of dust \citep{1998Kennicutt}.
To correct for dust, a nebula is assumed to be optically thick to ionising photons, using the known line-intensity ratio of H~$\alpha$ and H~$\beta$ of 2.86, as dictated by quantum mechanics \citep{2006Osterbrock}.
For this a Case B recombination is assumed, in which the electron temperature ($T_{e}$) is $T_{e}=10\,000~\rm{K}$ and the electron density ($n_{e}$) is $n_{e}=100~\rm{cm}^{-3}$ (standard for star-forming galaxies).
Applying the \citet{1994ODonnell} extinction law (with the extinction coefficients $\rm{k}_{\alpha}=2.52$ and $\rm{k}_{\beta}=3.66$), it is possible to derive the total visual extinction ($A_{V}$).
This approach works reasonably well for local star-forming galaxies, resulting in a range of $A_{V}=$1--2$~\rm{mag}$ \citep{1983Kennicutt,1998Kennicutt,2009Kennicutt}.

% To compute $\Sigma_{\mathrm{SFR}}$ from the luminosity of H~$\alpha$ corrected for internal extinction ($L_{\mathrm{H~\alpha,cor}}$; assuming H~$\alpha/$H~$\beta=2.86$), we follow the \cite{2013Calzetti} prescription as
To compute $\Sigma_{\mathrm{SFR}}$ from the luminosity of H~$\alpha$ corrected for internal extinction ($L_{\mathrm{H~\alpha,cor}}$), we follow the \cite{2013Calzetti} prescription as

\begin{equation}\label{eq:eq1}
    \Sigma_{\mathrm{SFR}} = 5.5 \times 10^{-42}\,L_{\mathrm{H~\alpha,cor}}\,\cos i,
\end{equation}

\noindent where the factor $5.5 \times 10^{-42}$ is in units of $\frac{\mathrm{M}_{\odot}\,\mathrm{yr}^{-1}}{\mathrm{erg}\,\mathrm{s}^{-1}}$ and the H~$\alpha$-SFR conversion uses a \cite{2001Kroupa} initial mass function (IMF).

ALMA observations of CO(2--1) are used to trace the molecular gas, at millimetre wavelengths unaffected by dust extinction.
To calculate $\Sigma_{\rm{mol}}$, we follow the \cite{2024Schinnerer} prescription, which accounts for the fraction of the CO-dark molecular gas, the emissivity, and the opacity of CO-emitting regions as

\begin{equation}\label{eq:eq2}
    \Sigma_{\mathrm{mol}} = \alpha_{\mathrm{CO,MW}}^{1-0}\,\left( Z / Z_{\odot}\right)^{-1.5}\,R_{21}^{-1}\,I_{\mathrm{CO(2-1)}} \cos i,
\end{equation}

\noindent where $\alpha_{\mathrm{CO,MW}}^{1-0}=4.35~\mathrm{M}_{\odot}~\left( \mathrm{K}\,\mathrm{km}\,\mathrm{s}^{-1}\,\mathrm{pc}^{2} \right) ^{-1}$ is the CO-to-H$_{2}$ conversion factor assuming a Milky Way calibration \citep{2013Bolatto}, $Z$ is the metallicity, $R_{21}=0.65\times\frac{\Sigma_{\mathrm{SFR}}}{1.8\times10^{-2}}~\left( \mathrm{M_{\odot}}\,\mathrm{yr}^{-1}\,\mathrm{kpc}^{-2} \right)^{-1} $ is the CO(2--1)-to-CO(1--0) line ratio as a function of $\Sigma_{\rm SFR}$ from local ISM conditions \citep{2025denBrok}, and $I_{\mathrm{CO(2-1)}}$ is the line-integrated CO(2--1) intensity in units of K\,km\,s$^{-1}$.
The metallicity $Z$ is converted from the oxygen-abundances in the nebular regions (Appendix~\ref{app:app1}).
In cases of non-detections of molecular gas surface density (for $\Sigma_{\rm{mol}}<2\sigma$), $2\sigma$ upper limits were used.

For CCSNe, we extracted the environmental parameters using the coarsest resolution available from VLT/MUSE and ALMA.
If VLT/MUSE had the coarsest resolution, ALMA data
were convolved with a Gaussian with the VLT/MUSE full width at half maximum.
Otherwise, VLT/MUSE observations were smoothed to the ALMA beam shape (with the respective major and minor axes).
SFE (or $\tau_{\mathrm{dep}}$) is computed as the ratio of $\Sigma_{\rm{SFR}}$ and $\Sigma_{\rm{mol}}$ as

\begin{equation}\label{eq:eq3}
    \mathrm{SFE}= \tau_{\mathrm{dep}}^{-1}=\Sigma_{\rm{SFR}}/\Sigma_{\rm{mol}}.
\end{equation}

The SFE is defined as the rate at which available molecular gas mass is converted into stars.
However, gas in these regions can be ionised due to various sources.
Examples of these sources are active galactic nucleus (AGNs) and massive stars.
Therefore, to select regions of star formation where the gas is primarily excited by massive stars, we performed an emission-line diagnostic.
These regions are identified using the so-called Baldwin, Phillips \& Terlevich (BPT) diagram \citep{1981Baldwin}, as detailed in Fig.~\ref{fig:figA1} (Appendix~\ref{app:app2}).
Since regions within galaxies can be ionised due to AGNs, increasing the observed value of H~$\alpha$, we removed 20~per cent of AGN-dominated galaxy pixels to obtain a clean sample.
Similarly, we removed two CCSNe (SN~2019ehk in NGC~4321 and SN~2023dpj in NGC~5135).
Their host galaxies contain AGNs \citep{2005GarciaBurillo,2018Sabatini}, and their galactocentric distances are 1.8 and 1.1~kpc, respectively.
After this correction, the final sample is 21 H-rich SNe and 19 SESNe, exploding in 25 host galaxies.

% Since regions can be ionised due to active galactic nuclei (AGN), an emission-line diagnostic was performed in order to select gas excited from massive stars.
% Hence, to obtain a clean sample, we removed two CCSNe (SN~2019ehk in NGC~4321 and SN~2023dpj in NGC~5135) and 20 per cent of galaxy pixels located in regions not dominated by \ion{H}{ii} emission.
% We identified these regions using the so-called Baldwin, Phillips \& Terlevich (BPT) diagram \citep{1981Baldwin}, as detailed in Fig.~\ref{fig:figA1} (Appendix~\ref{app:app2}).
% After this correction, the final sample results in 21 H-rich SNe and 19 SESNe, exploding in 25 host galaxies.

To compare these measurements with the properties of the host galaxies, we also repeated the procedure for random PHANGS pixels.
To give each galaxy a similar statistical weight, we randomly selected 10\,000 pixels per galaxy.
This gives us a reference estimate of the PHANGS's average properties of \ion{H}{ii} regions within galaxies (hereafter referred to as host galaxies).

%---RESULTS---
\section{Results}\label{sec:sec3}

\begin{figure*}
    \centering
    \resizebox{\hsize}{!}{\includegraphics{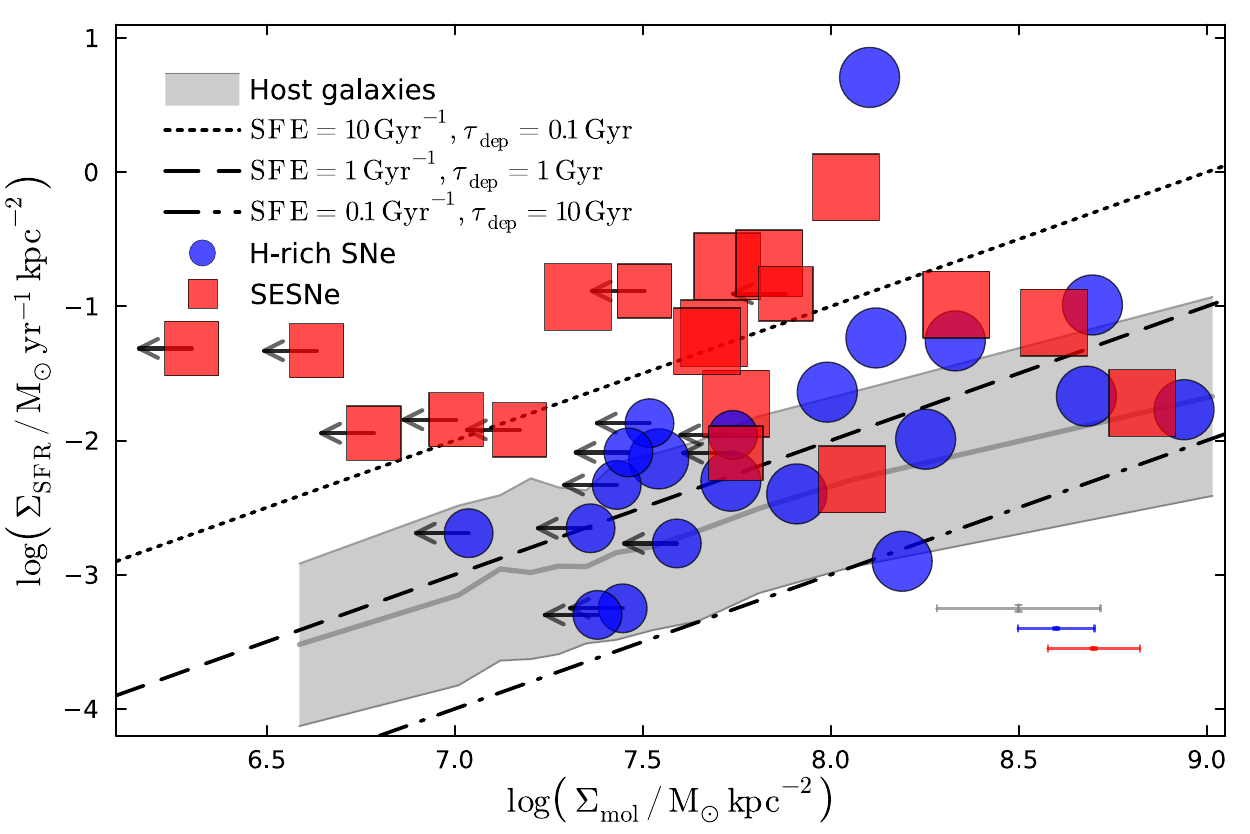}}
    \caption{\textbf{The star-formation law for the locations of CCSNe and host galaxies.}
    H-rich SNe and SESNe are denoted as blue circles and red squares, respectively.
    Black arrows represent 2$\sigma$ upper limits in cases of non-detections of $\Sigma_{\rm{mol}}$ (and symbols are smaller in comparison with detections).
    % The grey shaded region is the 16, 50, and 84 per cent scatter of host galaxy pixels (only detections are shown here).
    The grey shaded regions shows the scatter between 16th and 84th percentile of the detections in host galaxies, with the 50th percentile shown as a grey line.
    Also, different cases of star formation efficiencies (or molecular gas depletion times) are included for $10~\mathrm{Gyr}^{-1}$ ($0.1~\rm{Gyr}$), $1~\mathrm{Gyr}^{-1}$ ($1~\rm{Gyr}$), and $0.1~\mathrm{Gyr}^{-1}$ ($10~\rm{Gyr}$), represented by a dotted, dashed, and dash-dotted black line, respectively.
    The grey, blue, and red error bars depict the typical uncertainties for host galaxies, H-rich SNe, and SESNe, respectively.
    }
    \label{fig:fig1}
\end{figure*}

The star-formation law \citep[or Kennicutt--Schmidt scaling relation;][]{1959Schmidt,1998Kennicutt} is an empirical relation of SFR and gas density within galaxies.
Figure~\ref{fig:fig1} presents the star-formation law (only for molecular gas content) for the locations of our CCSNe.
At the explosion sites of all CCSNe, H~$\alpha$ emission was detected in the VLT/MUSE observations.
CO(2--1) was detected at the positions of 23 CCSNe (12 H-rich SNe and 11 SESNe), and 17 were not detected (9 H-rich SNe and 8 SESNe).
We also include the pixel distributions of host galaxies as a comparison sample ($\sim$\,20 per cent of CO host pixels are non-detections).
The constant SFE relations ($\tau_{\mathrm{dep}}$) are shown for $10~\mathrm{Gyr}^{-1}$ ($0.1~\mathrm{Gyr}$), $1~\mathrm{Gyr}^{-1}$ ($1~\mathrm{Gyr}$), and $0.1~\mathrm{Gyr}^{-1}$ ($10~\mathrm{Gyr}$).
This figure shows that H-rich SNe lie closer to the lower-SFE (longer depletion times) values, while SESNe lie closer to the higher-SFE (shorter depletion times) values.

\begin{figure*}
    \centering
    \resizebox{\hsize}{!}{\includegraphics{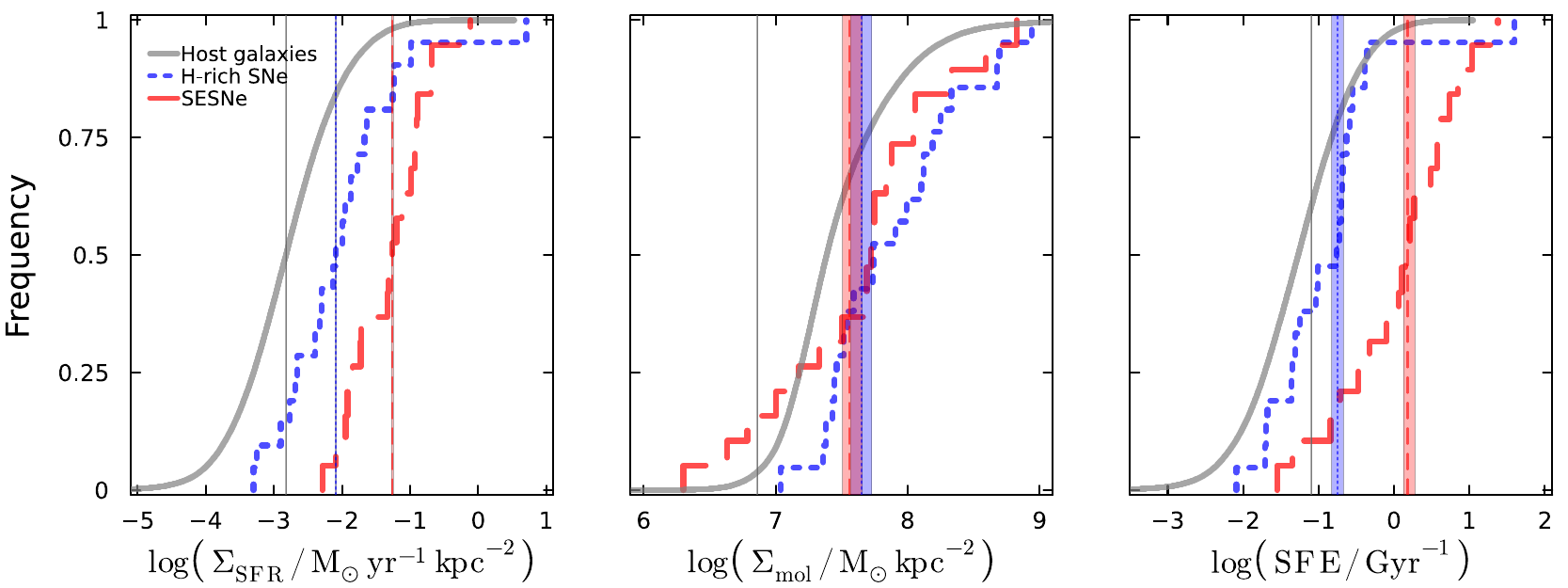}}
    \caption{\textbf{Warm ionised gas and molecular gas ISM conditions for CCSNe and host galaxies:} star formation rate surface density (\textit{left panel}), molecular gas surface density (\textit{middle panel}), and star formation efficiency (\textit{right panel}).
    For the case of non-detections, 2$\sigma$ upper limits are used.
    eCDFs of host galaxies, H-rich SNe, and SESNe, are shown as grey solid lines, blue dotted, and red dashed, respectively.
    The grey solid, blue dotted, and red dashed vertical lines represent the $1\sigma$ confidence intervals (including non-detections) of host galaxies, H-rich SNe, and SESNe, respectively.
    }
    \label{fig:fig2}
\end{figure*}

Figure~\ref{fig:fig2} shows empirical cumulative distribution functions (eCDFs) for CCSNe and host galaxies in star formation rate, molecular gas mass, and star formation efficiency.
Their 1$\sigma$ confidence intervals (computed using 1000 Monte Carlo simulations) are shown in Table~\ref{tab:tab1}.
To test whether distributions differ significantly (p-value $< 0.05$), Table~\ref{tab:tab2} reports two-sample Kolmogorov--Smirnov (KS) and Anderson--Darling (AD) tests.
A summary of the properties of the host-galaxy and the results derived is provided in Table~\ref{tab:tabC1} and Table~\ref{tab:tabC2} (Appendix~\ref{app:app3}), respectively.
Our results are also available in the online Supplementary Material.

\begin{table*}
    \centering
    \caption{Medians and 1$\sigma$ confidence intervals of star formation rate, molecular gas density, and star formation efficiency for each sample.}
    \label{tab:tab1}
    \centering
    \begin{tabular}{l|ccc}
    \hline
    Parameter & $\log \left( \Sigma_{\rm{SFR}} \, / \,  \mathrm{M}_{\odot}\,\mathrm{yr}^{-1}\,\mathrm{kpc}^{-2}  \right) $ & $\log  \left( \Sigma_{\rm{mol}} \, / \,   \mathrm{M}_{\odot}\,\mathrm{kpc}^{-2}  \right)$ & $\log \left( \rm{SFE} \, / \, \mathrm{Gyr}^{-1}  \right) $\\
    \hline
    Host galaxies      & $-2.8227^{+0.0002}_{-0.0002}$ & $6.86150^{+0.00003}_{-0.00001}$ & $-1.0954^{+0.0003}_{-0.0003}$ \\
    H-rich SNe  & $-2.09^{+0.01}_{-0.01}$    & $7.7^{+0.1}_{-0.1}$          & $-0.74^{+0.09}_{-0.08}$       \\
    SESNe       & $-1.26^{+0.01}_{-0.01}$       & $7.6^{+0.1}_{-0.1}$          & $0.18^{+0.06}_{-0.10}$        \\
    \hline
    \end{tabular}
\end{table*}

\begin{table*}
    \centering
    \caption{Two-sample KS and AD tests for host galaxies, H-rich SNe, and SESNe (star formation rate surface density, molecular gas surface density, and star formation efficiency).}
    \label{tab:tab2}
    \centering
    \begin{tabular}{l|ccccc}
        \hline
        Two-sample test & KS statistic & KS p-value & AD statistic & AD p-value & Null hypothesis rejected?\\
        {}              & {}            & {}          & {}            & {}          & (p-value\,$<0.05$)\\
        \hline
        $\Sigma_{\rm{SFR}}  \rightarrow$ Host galaxies \,\, vs. H-rich SNe & 0.4 & $6 \times 10^{-4}$  & 10 & $8 \times 10^{-6} $ & \textbf{Yes}\\
        $\Sigma_{\rm{SFR}}  \rightarrow$ Host galaxies \,\, vs. SESNe      & 0.8 & $1 \times 10^{-10}$ & 40 & $9 \times 10^{-20}$ & \textbf{Yes}\\
        $\Sigma_{\rm{SFR}}  \rightarrow$ H-rich SNe vs. SESNe       & 0.5 & $1 \times 10^{-2}$  & 6  & $1 \times 10^{-3} $ & \textbf{Yes}\\
        $\Sigma_{\rm{mol}}\,\rightarrow$ Host galaxies \,\, vs. H-rich SNe & 0.5 & $5 \times 10^{-4}$  & 10 & $2 \times 10^{-5} $ & \textbf{Yes}\\
        $\Sigma_{\rm{mol}}\,\rightarrow$ Host galaxies \,\, vs. SESNe      & 0.4 & $7 \times 10^{-3}$  & 4  & $8 \times 10^{-3} $ & \textbf{Yes}\\
        $\Sigma_{\rm{mol}}\,\rightarrow$ H-rich SNe vs. SESNe       & 0.3 & $5 \times 10^{-1}$  & 1  & $3 \times 10^{-1} $ & No          \\
        $\mathrm{SFE}\,\,   \rightarrow$ Host galaxies \,\, vs. H-rich SNe & 0.3 & $4 \times 10^{-2}$  & 3  & $2 \times 10^{-2} $ & \textbf{Yes}\\
        $\mathrm{SFE}\,\,   \rightarrow$ Host galaxies \,\, vs. SESNe      & 0.7 & $6 \times 10^{-8}$  & 40 & $8 \times 10^{-20}$ & \textbf{Yes}\\
        $\mathrm{SFE}\,\,   \rightarrow$ H-rich SNe vs. SESNe       & 0.7 & $1 \times 10^{-4}$  & 7  & $2 \times 10^{-4} $ & \textbf{Yes}\\
        \hline
    \end{tabular}
\end{table*}

For the measured star formation rate density (in units of $\mathrm{M}_{\odot}\,\mathrm{yr}^{-1}\,\mathrm{kpc}^{-2}$), we find a clear increasing trend from: host galaxy pixels ($\log \Sigma_{\mathrm{SFR}} = -2.8227^{+0.0002}_{-0.0002}$), to H-rich SNe ($\log \Sigma_{\mathrm{SFR}} = -2.09^{+0.01}_{-0.01}$), and SESNe ($\log \Sigma_{\mathrm{SFR}} = -1.26^{+0.01}_{-0.01}$). 
This is also supported by the two-sample tests of KS and AD as the p-values are significantly lower than 0.05, so the null hypothesis is rejected and the distributions are drawn from different parent populations.
In summary, SESNe preferentially explode in regions of higher SFR than H-rich SNe, and both CCSN types lie above the host-galaxy baseline.

For the computed molecular gas surface density (in units of $\mathrm{M}_{\odot}\,\mathrm{kpc}^{-2}$), a similar analysis yields different results.
Both CCSN classes occur in regions with higher molecular-gas surface density ($\log \Sigma_{\mathrm{mol}}= 7.7^{+0.1}_{-0.1}$ for H-rich SNe and $\log \Sigma_{\mathrm{mol}} = 7.6^{+0.1}_{-0.1}$ for SESNe) than the sample of host galaxy pixels ($\log \Sigma_{\mathrm{mol}} = 6.86150^{+0.00003}_{-0.00001}$.
The $\Sigma_{\mathrm{mol}}$ in the environments of H-rich SNe and SESNe are consistent within uncertainties.
This is also supported by the high KS and AD p-values (0.5 and 0.3, respectively).
This indicates that the observed difference between H-rich SNe and SESNe is not statistically significant, so the null hypothesis cannot be rejected.
On the other hand, a significant difference is found when $\Sigma_{\mathrm{mol}}$ in the environments of both CCSNe are compared with the host galaxy pixels.
This is also confirmed by low p-values for the two-sample tests of KS and AD, comparing the host galaxies with H-rich SNe (KS $\mathrm{p}$-$\mathrm{value}$$\,\sim$\,$5\times10^{-4}$ and AD $\mathrm{p}$-$\mathrm{value}$$\,\sim$\,$2\times10^{-5}$) and SESNe ($\sim$\,$7\times10^{-3}$ and $\sim$\,$8\times10^{-3}$).
Altogether, both CCSNe explode in regions with similar molecular gas surface densities, but at densities higher than the host galaxy average.

When the SFE is analysed, a similar trend is found compared to the results of SFR.
The SFE values (in units of Gyr$^{-1}$) are as follows: host galaxy pixels ($\log \rm{SFE}=-1.0954^{+0.0003}_{-0.0003}$) $<$ H-rich SNe ($\log \rm{SFE}=-0.74^{+0.09}_{-0.08}$) $<$ SESNe ($\log \rm{SFE}=0.18^{+0.06}_{-0.10}$).
The distributions of host galaxies with H-rich SNe give p-values of $0.04$ for KS and $0.02$ for AD, respectively.
Also, for SFE of host galaxies and SESNe, the p-values are $6\times10^{-8}$ for KS and $8\times10^{-20}$ for AD, respectively.
A clear difference in star formation efficiency values is also found when comparing H-rich SNe and SESNe.
This is supported by the low p-values of $1\times10^{-4}$ and $2\times10^{-4}$ for KS and AD, respectively.
Although we work with values of $\Sigma_{\mathrm{mol}}$ corrected by $12\,+\, \left( \mathrm{O}/\mathrm{H} \right)$, our trends do not change whether this metallicity is taken into account or not.

%---DISCUSSION---
\section{Discussion}\label{sec:sec4}

First, the main outcome of this work is that SFE values increase in the following order: host galaxies $\rightarrow$ H-rich SNe $\rightarrow$ SESNe.
Second, the SFR trends are consistent with previous studies, with higher values for SESNe than for H-rich SNe \citep{2008Anderson,2012Anderson,2013Crowther,2014Galbany,2018Galbany,2018Kuncarayakti,2023Pessi}, and both CCSN types being above the host-galaxy average \citep{2024MaykerChen}.
Third, our molecular gas values show trends similar to \cite{2024Solar} -- 
%molecular gas mass:\,
$\mathrm{host}$ $\mathrm{galaxies} < \mathrm{H}$-$\mathrm{rich}$ $\mathrm{SNe}$\,$\sim$\,$\mathrm{SESNe}$ -- but different from \cite{2017Galbany} and \cite{2023MaykerChen} -- %Molecular gas mass:\, 
$\mathrm{H}$-$\mathrm{rich}$ $\mathrm{SNe} < \mathrm{SESNe}$.
% Because previous studies have already investigated the star formation rate and molecular gas environments separately, we focus here on the physical interpretation of our star formation efficiency results.

In summary, we find similar $\Sigma_{\mathrm{mol}}$ environments for SESNe and H-rich SNe, but higher $\Sigma_{\mathrm{SFR}}$ for the former.
There is roughly an order-of-magnitude ($\times 8$) difference in SFE values between SESNe and H-rich SNe.
This can be explained by two progenitor scenarios of high-mass stars ($\geq 8~\mathrm{M}_{\odot}$) for SESNe, both differing from the H-rich SNe: very massive stars \citep[$\geq20$~M$_{\mathrm{\odot}}$;][]{1993Woosley} or binary systems \citep[$\leq20~\mathrm{M}_{\mathrm{\odot}}$;][]{1992Podsiadlowski}.
For the very-massive-star case, the outer layers are removed due to strong mass-loss (given the high initial stellar mass).
% We show below that this is consistent with the data only if the IMF in these regions is top-heavy.
We show below that this could be consistent with the data if the IMF in these regions is top-heavy.
For the latter case, a companion is responsible for stripping away the envelope, and this mechanism is more effective at lower initial masses than for very massive stars.
This can explain our data because, as we discuss below, higher binarity fraction (possibly boosted by increased turbulence) implies higher H~$\alpha$ luminosity.
% In both scenarios, SESN progenitors are more massive and/or have a higher binarity fraction than H-rich SNe.}

\subsection{Very Massive Stars}

The H~$\alpha$ emission is expected to be linked to massive stars \citep[][and references therein]{2012Kennicutt}.
This is because the most massive stars are responsible for ionising the surrounding gas medium through intense UV radiation.
We find that the SFR at the explosion sites of both CCSN populations are higher compared to the sample of all host-galaxy pixels.
Many of the galaxy pixels are not related to young and intense star-forming regions, therefore the host galaxy distribution will be shifted towards a lower SFR.

Here we shed light on the outcome that SESNe explode in regions of higher SFR than those of H-rich SNe, implying a possible difference in their progenitor properties.
For instance, SESN progenitors could be more massive than H-rich SNe, and therefore their lifetimes are shorter, which explains higher SFRs \citep[see also][]{2026Xi}.
In other words, the explosion delay time after the formations of progenitors of more massive stars is shorter, so their parents clouds should have a high SFR, whereas for lower-mass stars at the time of the explosion the SFR has already decreased.

In the very-massive-star scenario, SESNe are associated with higher-SFR regions because they host higher numbers of such stars.
% The environmental properties may imply that SESN populations are biased toward more massive stars.
For a fixed amount of molecular gas, this could produce a higher flux of H~$\alpha$ per unit of gas.
% , which would produce a higher SFR.
% Because the SFE traces the current state of the region, a higher SFE may indicate shorter depletion times (i.e., more efficient ongoing star formation).
Because higher star formation rates correspond to younger, short-lived stellar populations, they can also increase the number of single-star SN progenitors.
% However, this is not consistent with our results when the amount of molecular gas is included in the analysis.

However, once molecular-gas densities at their positions are included through this multiwavelength approach, a different interpretation of CCSN progenitors emerges, as we present in what follows.
The amount of gas present at the time of the explosion dissipate over time due to feedback from young stars \citep[such as stellar winds, photoionisation, radiation pressure, and SNe;][]{1978Spitzer}.
In other words, the molecular gas mass at the CCSN positions can be used as an indicator of their progenitor lifetimes \citep{2024Solar}.
More massive stars (associated with higher H~$\alpha$ emission) would be expected to explode in the densest regions of the molecular gas.
However, this is inconsistent with the fact that we observe similar values of molecular gas surface density at the locations of SESNe and H-rich SNe which implies that their progenitor lifetimes and therefore initial masses are similar \citep{1992Schaller}.

The higher SFE at the locations of SESNe than for H-rich SNe but similar molecular gas densities could alternatively be explained by variations in the IMF.
SESN progenitors could be preferentially formed in top-heavy IMF regions \citep{2018Schneider}.
This means that the regions hosting SESNe can be similar to those hosting H-rich SNe with regards to all properties, but the number of very massive stars for the former.
This would leave the molecular gas distribution similar for both types of CCSNe, but those hosting SESNe would have higher H~$\alpha$ luminosities due to higher number of very massive stars.
This is consistent with our measurements.

\subsection{Binary Systems}

We also suggest another scenario for the progenitor channels of SESNe in which they arise from binary systems.
%This could be explained via binary channels for SESN progenitors, as 
This is then consistent with similar molecular gas masses at their positions and at those of H-rich SNe, because the binary systems are expected to have lifetimes similar to those of H-rich SN progenitors.

Growing evidence supports the binary model for SESNe, such as simulations of helium stars \citep{2020Dessart}, emission-line diagnostics of stellar populations \citep{2019Xiao}, periodic undulations evidenced from a bound compact remnant \citep{2023Moore,2024Chen,2026Cartier}, high relative rate of SESNe and H-rich SNe \citep{2011Smith,2017Shivvers}, direct detections of binary companions at the SN site \citep{2004Maund,2015Maund,2022Fox,2026Zapartas}, using molecular gas as tracer of progenitor lifetimes \citep{2024Solar}, and low ejecta masses \citep{2011Drout,2016Modjaz}.
Lightcurve studies also suggest this with modelling \citep{2016Lyman,2021Barbarino} and early-excess lightcurve characteristics \citep{2024Das,2025Ayala,2025Chiba}.
Moreover, there is a lack of evidence for Wolf-Rayet stars (or luminous progenitors) for most of SESNe \citep{2013Eldridge,2015Smartt}: difficulty to detect massive stars as progenitors \citep{2012Yoon,2013Cao,2018Kilpatrick,2018VanDyk} and failed explosions \citep[i.e., the massive star collapses into a black hole producing an undetectable transient;][]{2011OConnor,2016Ertl,2020Patton,2021bZapartas}.
% stellar evolution models suggest failed explosions in WR stars \citep[i.e., the SESN collapses into a black hole with a faint transient;][]{2021bZapartas}, difficulty to detect massive WR stars as progenitors \citep{2013Cao,2018VanDyk,2018Kilpatrick}, and significant possibility for massive WR stars to implode \citep{2011OConnor,2016Ertl,2020Patton,2021bZapartas} potentially with no transients \citep{2026De}.
% The binary system scenario for SESN progenitors is supported due to the overwhelming evidence stated above.

It is essential to include binary evolution when modelling synthetic stellar spectra in galaxies.
Specifically, when binary star models are included, they boost the ionisation of \ion{He}{II} (at $1640~\mathrm{\mathring{A}}$) emission lines \citep{2012Eldridge,2017Eldridge}, cosmic reionisation of hydrogen \citep{2020Gotber}, \ion{H}{ii} regions \citep{2025CournoyerCloutier}, and a number of young, metal-poor, actively star-forming galaxies \citep{2024Lecroq}.
Similarly, younger open clusters tend to exhibit a higher binary fraction than older counterparts \citep{2025Alexander}, consistent with more recent star formation and higher H~$\alpha$ emission.
In summary, pre-SN feedback from binary stars (mechanical or radiative) enhances the overall ionising radiation from young stellar populations.
These effects may explain why SESNe are located in regions with higher H~$\alpha$ luminosities that those for H-rich SNe, despite having similar molecular gas properties.

% Binary interactions are considered crucial for explaining many of the properties of the majority SESNe.
% Massive stars, to which all CCSN progenitors belong, are observed to have very high interacting binary fractions \citep[$>\,$$70$ per cent of O-type stars;][]{2012Sana,2017Moe}.
% The single-star channel for SESNe requires very massive stars, and those are rare because their formation rate scales roughly linearly with the star formation rate.
% However, we do not rule out bimodal distribution of the mechanisms responsible for removing the outer layers of SESNe.
% Moreover, SESN progenitors could arise from a population with slightly higher initial masses than H-rich SN progenitors and also a higher binarity fraction (with a higher incidence of interacting systems).

Concentrating now on the H-rich SNe, their progenitors arise from high-mass stars, and our results for $\Sigma_{\mathrm{SFR}}$ and $\Sigma_{\mathrm{mol}}$ are consistent with this interpretation when compared to random pixels from the host galaxies (both with six times higher values than typical pixels).
This is also consistent with the SFE being around two times greater for H-rich SNe than for host galaxies.
Given this difference in SFE, there is a clear trend for progenitor channels of H-rich SNe to be associated with high-mass stars.
H-rich SNe are also expected to be produced by a significant fraction of binary systems ($\sim$\,$30$--$50$ per cent), mostly from mass accretion or merging processes \citep{2019Zapartas}.
The binary progenitors of H-rich SNe explode later than the red supergiants from single stars of the same final luminosity \citep[i.e., core mass;][]{2021aZapartas}.
% The main reason is that lower initial binary masses imply longer evolutionary timescales, so the interaction occurs later.
Theoretically, this prolonged delay-time can even exceed the lifetime of all single star CCSN progenitors, going into the regime of 50--200 Myr, when all single stars have already exploded \citep[$\leq50~$Myr;][]{2017Zapartas}.
However, the more pronounced difference in SFE values for SESNe and host galaxies (around 20 times greater) could reflect more extreme conditions of their progenitors.
This may imply that more efficient star-forming regions enhance the production of interacting binaries and the multiplicity fraction.

The related interpretation is that regions hosting SESNe are preferentially those above the Kennicutt–Schmidt scaling relation with higher $\Sigma_{\mathrm{SFR}}$ for their $\Sigma_{\mathrm{mol}}$ have increased binary fraction, possibly due to turbulences.
GMCs are short-lived structures ($\sim$10--30~Myr) where star formation proceeds fast and inefficiently since only a small fraction of the molecular gas is converted into stars \citep{2017Semenov,2019Kruijssen,2020Chevance,2022Kim}.
The GMCs that lie above the scatter of the star-forming scaling relation represent regions with an unusually high SFR for their given $M_{\rm mol}$ (boosted SFE and shortened $\tau_{\rm dep}$).
One physical reason for the scatter comes from the supersonic, kinetic (non-thermal) molecular gas turbulence.
Turbulence plays a dual role in star formation: supports molecular clouds against global gravitational collapse (low SFE) and creates shock waves that compress gas locally, forming dense filaments and clumps, which are the seeds for new stars (high SFE).
CCSNe (H-rich SNe and SESNe) are found in turbulent GMCs \citep{2026Solar}, suggesting that their progenitors have increased formation in high densities and/or are associated with a binary nature.
Similarly, the binarity fraction increases with the initial stellar mass \citep{2017Moe,2023Offner}, which is also directly related to the SFR.

We do not rule out bimodal distribution of the mechanisms responsible for removing the outer layers of SESNe.
Moreover, SESN progenitors could arise from a population with slightly higher initial masses than H-rich SN progenitors and also a higher binarity fraction (with a
higher incidence of interacting systems).

\subsection{Caveats}

% One caveat in our study is the ratio of different types of CCSN.
Using a median redshift of $z=0.0149$ (distance $\sim$\,$60$~Mpc), \cite{2025Pessi} find volumetric fractions of all CCSNe of $\sim$\,$60$ and $\sim$\,$30$ per cent for H-rich SNe and SESNe \citep[see also][]{2026Ercolino}, respectively.
Also, from this work, the specific (per unit stellar mass) CCSN rate increases for galaxies with lower stellar mass.
In our case, the ratio is different, as we have 21 H-rich SNe and 19 SESNe, and the sample is composed only of massive host galaxies ($\sim$\,$\log \left[ 10.5 / \mathrm{M}_{\odot}\right]$).
Our host galaxies are not a homogeneous sample and are biased toward large and massive galaxies, which makes it systematically inclined to metal-rich galaxies \citep{2004ATremonti}.
A similar study done for an unbiased sample would resolve this tension, but would require significant investment of the ALMA observing time.

% Furthermore, our host galaxies are nearby targets ($<80$~Mpc), hence, the metallicities are higher in comparison with the earlier Universe.
% This might also affect our interpretation of the results as the fractions of CCSNe could differ.
Furthermore, our host galaxies are nearby targets ($<80$~Mpc), hence, the physics is constrained to conditions of the Local Universe.
Metallicities in the earlier Universe are lower in comparison with low-$z$ galaxies.
In our case, the metallicity correction $(Z/Z_{\odot})^{-1.5}$ from eq.~\ref{eq:eq1} resulted in a similar increase of 50 per cent for both CCSN locations and host galaxy pixels (for 0.75\,$Z_{\odot}$; see Appendix~\ref{app:app2}).
Conversely, for a metallicity of 0.25\,$Z_{\odot}$, the correction is almost an order of magnitude and this could affect the interpretation of results significantly (e.g., for high-$z$ galaxies).
That is why several proxies of the molecular gas should be used in addition (such as [C\,{\small I}] at both 370 and 609~$\mu$m and/or [C\,{\small II}] at 156~$\mu$m).
Further studies including other physical conditions (e.g., metal content, binarity fraction, turbulence, stellar populations, etc.) would be required to consider whether SFE alone is producing the observed outcomes.

SFR calibrators using H~$\alpha$ and H~$\beta$ breaks down for $A_{V} >2.5~\rm{mag}$ \citep{2009Kennicutt}.
In our case, typical galaxy pixels have $A_{V}=1$--$2$ mag values; 90~per cent of CCSN environments have $A_{V}<2.5~\rm{mag}$.
Only three CCSNe show high dust extinction values: 4.3, 2.9, and 4.6~mag for SN~1985P, SN~2011hp, and SN~2004ip, respectively.
However, when these highly extincted CCSNe are removed from the original sample, the trends stay similar.
$A_{V}$ is expected to increase in central regions of galaxies, so the observed H~$\alpha$ flux will be biased toward lower values.
This means that inner galactic regions could exhibit higher values of SFR and SFE than the those reported here.
In our data, since we compare CCSNe and host galaxy locations at similar galactocentric distances, this effect would not alter the results.
% In our data, this makes our conclusions even stronger as SFR and SFE regions of host galaxies are directly compared to SESN and H-rich SN explosion sites.
% Furthermore, since we compare to pixel values at similar galactocentric distances, this effect would not alter the results.

Although we refer to cloud-scale observations at $\sim$100~pc spatial scales, the actual \ion{H}{ii} region and GMC sizes span the range from a few tens of parsecs up to approximately 100~pc \citep{2010Fukui,2011Lopez}.
Due to instrumental limitations, increasing the spatial resolution would reduce the sample size.
For this reason, to achieve the goals of mapping individual structures in the ISM, we used a homogenised sample from VLT/MUSE and ALMA observations, leveraging the capabilities of these telescopes as the best in the world in their respective wavelength regimes.

% Another caveat concerns the existence of mergers and gas disturbances.
% The ISM properties provide information on the conditions under which the CCSN progenitors are produced.
% In particular, molecular gas is the main driver that impacts star formation activity \citep[][and references therein]{2022Saintonge}.
% Star formation is quenched when the molecular gas is stabilised against collapse and unable to form stars.
% The typical molecular gas depletion time in spiral galaxies in the Local Universe is about $1$~Gyr \citep{2011Saintonge,2013Tacconi}.
% However, galaxy mergers can enhance SFE by an order of magnitude, as can dynamical disturbances to a lesser degree (e.g., minor mergers, galaxies with companions, and galaxies with strong bars).

%---CONCLUSIONS---
\section{Conclusions}\label{sec:sec5}

We collected the surface densities of star formation rate and molecular gas for all available CCSNe to compute star formation efficiencies in their environments.
To do so, we used telescopes capable of resolving individual \ion{H}{ii} regions and GMCs ($\sim$\,100~pc): VLT/MUSE and ALMA, respectively.
A total of 21 H-rich SNe and 19 SESNe were collected (in 25 host galaxies).
Both populations of CCSNe explode in regions of higher SFE than the average of their host (PHANGS) galaxies, suggesting that progenitors are associated with massive stars.
We find that SESNe are preferentially located in regions with significantly higher star formation rate surface densities, but comparable molecular gas surface densities relative to H-rich SNe.
This results in star formation efficiencies (molecular gas depletion times) that are higher (shorter) by an order of magnitude at SESN locations than at H-rich SN.
In summary, SESNe preferentially explode in environments characterised by highly efficient, intense star formation.
We conclude that the observed properties are best explained by the following scenarios (not mutually exclusive):

 \begin{enumerate}
     \item Top-heavy IMF: regions hosting SESN explosions may be biased toward a top-heavy IMF.
     In this case, an excess of very massive stars boosts the observed H~$\alpha$ luminosity (and thus the measured SFR) without a corresponding increase in the total molecular gas reservoir.
     \item Binaries dominate the overall ionising radiation: if SESN progenitors predominantly arise from binary systems, then their increased production of ionising photons could enhance the measured star formation rate at these locations.
     This allows for a high H~$\alpha$ flux for a given molecular gas density.
     \item Environmentally driven binarity: specific physical conditions (such as increased turbulence) favour a higher binary fraction.
     In this model, SESNe are tracers of efficient star-forming environments where binary interaction is the dominant evolutionary path.
 \end{enumerate}

% We discard the scenario where SESN progenitors originate only from very massive single stars in star-forming regions with normal IMFs.
It is unlikely the scenario where SESN progenitors originate only from very massive stars in star-forming regions with normal IMFs.
Such a model would imply that SESNe should be found in regions with higher $M_{\text{mol}}$ due to the shorter timescales involved, a prediction that is inconsistent with our data and previous works.
Ultimately, our results suggest that the \lq SFR–$M_{\text{mol}}$\rq~relation for CCSN environments is a powerful probe of progenitor physics, pointing towards either IMF variations and/or a high binary fraction as the primary driver for increased SESN rate.

This is the first time that the environments of SFE is computed for a significant sample of CCSNe.
Furthermore, these observations resolved cloud-scales for \ion{H}{ii} regions and GMCs.
Future telescopes such as ELT/HARMONI will have a strong impact in IFU studies at locations of SNe, as better resolutions (possible to match with ALMA), extended wavelengths coverage, and larger samples will be achieved.
In addition, to improve galaxy evolution models in numerical cosmological simulations, SESN populations should be implemented in the prescriptions, as they differ in the feedback injected into the ISM and chemical mixing from H-rich SNe.

\section*{Acknowledgements}

% This research was funded in whole or in part by the National Science Centre (NCN), Poland, project PRELUDIUM 2024/53/N/ST9/00350 and project OPUS 2023/49/B/ST9/00066.
This research was funded in whole or in part by the National Science Centre (NCN), Poland (grant numbers 2024/53/N/ST9/00350 and 2023/49/B/ST9/00066). 
For the purpose of Open Access, the author has applied a CC-BY public copyright licence to any Author Accepted Manuscript (AAM) version arising from this submission.
Supported by the Foundation for Polish Science (FNP).
J.N.~acknowledges the support of the Polish National Agency for Academic Exchange (NAWA) Bekker grant BPN/BEK/2023/1/00271, and the kind hospitality of the IAC.
L.G. acknowledges financial support from CSIC, MCIN and AEI 10.13039/501100011033 under projects PID2023-151307NB-I00, PIE 20215AT016, CEX2020-001058-M, and by the MaX-CSIC Excellence Award MaX4-SOMMA-ICE.
This work was supported by a research grant (VIL54489) from VILLUM FONDEN.
This research has made use of the services of the ESO Science Archive Facility.
Based on data products created from observations collected at the European Organisation for Astronomical Research in the Southern Hemisphere under ESO programme(s) 1100.B-0651 (PHANGS-MUSE; PI Schinnerer), 094.B-0321 (MAGNUM; PI Marconi), 0100.B-0116 (MAD; PI Carollo), 097.B-0640 (TIMER; PI Gadotti), 60.A-9194 (PI Kool), 095.D-0172 (AMUSING, PI Kuncarayakti), and 114.27G5.001 (PI Solar).
ALMA is a partnership of ESO (representing its member states), NSF (USA) and NINS (Japan), together with NRC (Canada), MOST and ASIAA (Taiwan), and KASI (Republic of Korea), in cooperation with the Republic of Chile.
The Joint ALMA Observatory is operated by ESO, AUI/NRAO and NAOJ.
The National Radio Astronomy Observatory is a facility of the National Science Foundation operated under cooperative agreement by Associated Universities, Inc.
This paper makes use of the following ALMA data:
ADS/JAO.ALMA\#2013.1.00243.S: PI Colina (archival), NGC~5135;
ADS/JAO.ALMA\#2013.1.01161.S: PI Sakamoto (PHANGS), NGC~1365;
ADS/JAO.ALMA\#2015.1.00902.S: PI Ao (archival), PGC~084885;
ADS/JAO.ALMA\#2015.1.00925.S: PI Blanc (PHANGS), NGC~1087, NGC~1566;
ADS/JAO.ALMA\#2015.1.00956.S: PI Leroy (PHANGS), NGC~1672, NGC~3627, NGC~4254, NGC~4303, NGC~4321, NGC~6744;
ADS/JAO.ALMA\#2017.1.00392.S: PI Blanc (PHANGS), NGC~1087, NGC~1433, NGC~1566;
ADS/JAO.ALMA\#2017.1.00886.L: PI Schinnerer (PHANGS), NGC~2997;
ADS/JAO.ALMA\#2018.A.00062.S: PI Faesi (PHANGS), NGC~5128;
ADS/JAO.ALMA\#2018.1.00272.S: PI Sliwa (archival), NGC~4038;
ADS/JAO.ALMA\#2018.1.01651.S: PI Leroy (PHANGS), NGC~1300, NGC~1087, NGC~1433, NGC~1566;
ADS/JAO.ALMA\#2019.1.01305.S: PI Tanaka (archival), NGC~4981;
ADS/JAO.ALMA\#2021.1.00099.S: PI Michałowski (ACOS survey), ESO~440-011, ESO~492-002, NGC~0908, NGC~1729, NGC~3278, NGC~3354, NGC~3810, NGC~4219, PGC~037625.
This research has made use of the Transient Name Server (TNS), operated by the IAU Supernova Working Group and hosted by the Weizmann Institute of Science.
We used IAU Circulars and CBETs presented by the Central Bureau for Astronomical Telegrams.
We acknowledge the use of the Supernova Catalog maintained by the Institute of Astronomy, Moscow State University (SAI MSU).
The catalog can be accessed at: http://stella.sai.msu.ru/sncat/.
We acknowledge the work of astronomers supporting the LEDA database (http://leda.univ-lyon1.fr/), the Asiago Supernova Catalogue (http://graspa.oapd.inaf.it/asnc.html), The STScI Digitized Sky Survey (http://archive.stsci.edu/cgi-bin/dss\_form), and CfA List of Supernovae (http://www.cbat.eps.harvard.edu/lists/RecentSupernovae.html).
This research has made use of the NASA/IPAC Extragalactic Database (NED), which is operated by the Jet Propulsion Laboratory, California Institute of Technology, under contract with the National Aeronautics and Space Administration (http://ned.ipac.caltech.edu/).
Facilities: ALMA; VLT: Yepun.
Software: \texttt{JULIA} \citep{2017Bezanson}, \texttt{STARLIGHT} \citep{2005CidFernandes}, \texttt{CARTA} \citep{2021Comrie}, \texttt{ASTROPY} \citep{2013Astropy,2022Astropy}, \texttt{NUMPY} \citep{2020Harris}, and \texttt{SCIPY} \citep{2020Virtanen}.
M.S. led the project, performed all the data analysis, and wrote the paper.
M.J.M. supervised the whole project and provided the ACOS data.
J.N. co-supervised the whole project.
L.G. provided the AMUSING data.
L.G., J.P.A., J.S., T.P., and E.Z. presented significant contributions to the interpretation of the data.
B.\v{S}., J.A., O.R., J.H., C.J., P.N., D.S., and A.L. contributed to the discussion of results and improve the text in the manuscript.
All co-authors provided input to the paper.

%%%%%%%%%%%%%%%%%%%%%%%%%%%%%%%%%%%%%%%%%%%%%%%%%%
\section*{Data Availability}

The data underlying this article are available in the article and in its online supplementary material.


\bibliographystyle{mnras}
\bibliography{biblio}

%%%%%%%%%%%%%%%%%%%%%%%%%%%%%%%%%%%%%%%%%%%%%%%%%%

%%%%%%%%%%%%%%%%% APPENDICES %%%%%%%%%%%%%%%%%%%%%

\appendix

\section{Metallicity corrections}\label{app:app1}

We correct the $\Sigma_{\rm{mol}}$ values from the metallicity.
To do so, we computed the gas-phase metallicity indicator described by \cite{2013Marino} as

\begin{equation}
    12 + \log \left( \mathrm{O}/\mathrm{H} \right) = 8.535 - 0.214 \times \mathrm{O3N2},
\end{equation}

\noindent where $\mathrm{O3N2} = \log \frac{[\ion{O}{III}] / \mathrm{H}~\beta}{\mathrm{[\ion{N}{II}}] / \mathrm{H}~\alpha}$, and \ion{N}{II} (at $6583.45~\mathrm{\mathring{A}}$) and \ion{O}{III} (at $5006.843~\mathrm{\mathring{A}}$) are the singly ionised nitrogen and doubly ionised oxygen, respectively.
Metallicity $Z$, which is the abundance of elements heavier than helium, was calculated in the following way:

\begin{equation}
   \left( Z/Z_{\odot} \right) = 
   10^{\left[12 + \left( \mathrm{O}/\mathrm{H} \right) \right]
   -
   \left[12 + \left( \mathrm{O}/\mathrm{H} \right) \right]_{\odot}},
\end{equation}

\noindent where according to \cite{2009Asplund}, $Z_{\odot} = 0.0142$ and $12 + \log \left( \mathrm{O}/\mathrm{H} \right)_{\odot} = 8.69$.
The factor $(Z/Z_{\odot})^{-1.5}$ used in the correction of $\Sigma_{\rm{mol}}$ is equal to $1.7451^{+0.0003}_{-0.0003}$, $1.62^{+0.04}_{-0.03}$, and $1.66^{+0.12}_{-0.04}$ for host galaxies, H-rich SNe, SESNe, respectively.

\section{Dominant ionisation source}\label{app:app2}

\begin{figure}
    \centering
    \resizebox{\hsize}{!}{\includegraphics{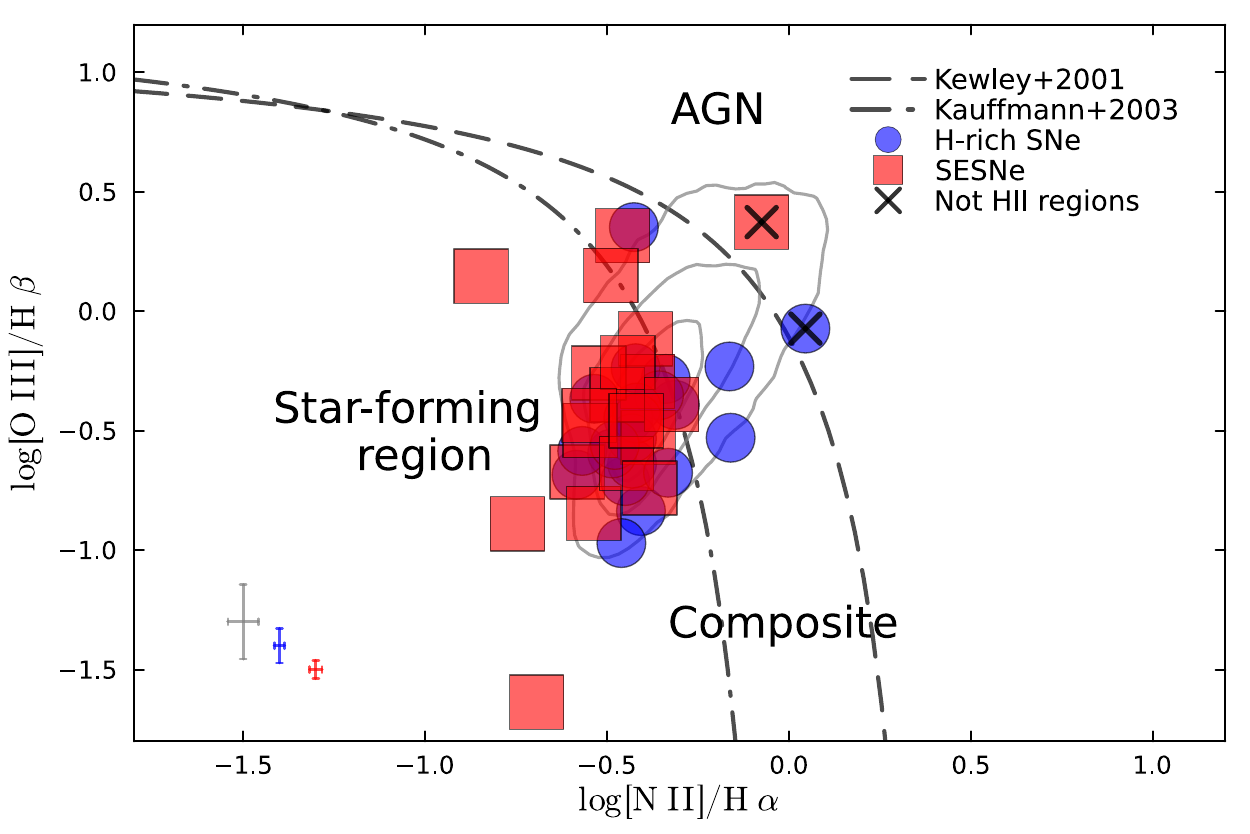}}
    \caption{\textbf{BPT diagram at locations of CCSNe and pixels of host galaxies.}
    H-rich SNe and SESNe are denoted as blue circles and red squares, respectively.
    The grey contours are the 25, 50, and 75 per cent representations of host galaxy pixels.
    Depending on the zone, the principal ionisation mechanism is labelled in the figure (star-forming region, AGN, and Composite).
    The source-separation curves from \citet{2003Kauffmann} and \citet{2001Kewley} are shown as black dash-dotted and dashed lines, respectively.
    For CCSNe, excluded targets are marked with black crosses.
    The grey, blue, and red error bars show the typical uncertainties for host galaxies, H-rich SNe, and SESNe, respectively.
    }
    \label{fig:figA1}
\end{figure}

Gas can be photoionised by star-forming regions or collisionally excited by AGNs \citep{2023Groves}.
The BPT diagnostic uses the ratios [\ion{O}{III}]/H~$\beta$ and [\ion{N}{II}]/H~$\alpha$ to pinpoint which is the dominant ionisation source (see Fig.~\ref{fig:figA1}).
We retained CCSNe located in star-forming and composite regions (a total of 40).
We excluded two CCSNe located in AGN regions because their environments are not primarily ionised by massive stars, which is required for calculating the star formation efficiency.
A total of 20 per cent of the pixels were removed from the BPT diagram for host galaxies.

\section{Host galaxies and supernova environments}\label{app:app3}

Detailed information on our host-galaxy sample and properties of the SNe is given in Table~\ref{tab:tabC1}.
This table lists the host-galaxy name, whether the observation was made by PHANGS, SN name, right ascension (R.A.), declination (Dec.), SN type, SN group, and redshift.

\begin{table*}
    \caption{CCSNe and their host galaxy properties.}
    \label{tab:tabC1}
    \centering
    \setlength{\tabcolsep}{3pt}
    \begin{tabular}{r|ccccccccc}
        \hline
        {} & Host galaxy & PHANGS- & PHANGS- & SN name & SN R.A. & SN Dec. & SN type & SN group & $z$\\
        {} & name & ALMA & VLT/MUSE & {} & {} & {} & {} & {} & {}\\
        {} & {}   & {}   &   {} & {} & {($\rm{h}\,\rm{min}\,\rm{s}$)} & {($\degr\,\arcmin\,\arcsec$)} & {} & {} & {}\\
        \hline
        1  & ESO~440-011 & No  & No  & LSQ13doo    & 11$^{\rm{h}}$48$^{\rm{m}}$45$\fs$780  & -28\degr17\arcmin31\farcs20 & Ic & SESNe & 0.006468 \\
        2  & ESO~492-002 & No  & No  & SN~2005lr    & 07$^{\rm{h}}$11$^{\rm{m}}$39$\fs$030  & -26\degr42\arcmin20\farcs20 & Ic & SESNe & 0.008647 \\
        3  & NGC~0908    & No  & No  & SN~1994ai    & 02$^{\rm{h}}$23$^{\rm{m}}$06$\fs$170  & -21\degr13\arcmin58\farcs30 & Ic & SESNe & 0.005026 \\
        4  & NGC~1087    & Yes & Yes & SN~1995V     & 02$^{\rm{h}}$46$^{\rm{m}}$26$\fs$770  & -00\degr29\arcmin55\farcs60 & II & H-rich SNe & 0.00507 \\
        5  & NGC~1300    & Yes & Yes & SN~2022acko  & 03$^{\rm{h}}$19$^{\rm{m}}$38$\fs$990  & -19\degr23\arcmin42\farcs68 & II & H-rich SNe & 0.005264 \\
        6  & NGC~1365    & Yes & Yes & SN~1983V     & 03$^{\rm{h}}$33$^{\rm{m}}$31$\fs$630  & -36\degr08\arcmin55\farcs00 & Ic & SESNe & 0.0055 \\
        7  & NGC~1365    & Yes & Yes & SN~2001du    & 03$^{\rm{h}}$33$^{\rm{m}}$29$\fs$110  & -36\degr08\arcmin32\farcs50 & II & H-rich SNe & 0.00371 \\
        8  & NGC~1433    & Yes & Yes & SN~1985P     & 03$^{\rm{h}}$42$^{\rm{m}}$06$\fs$300  & -47\degr12\arcmin35\farcs60 & II & H-rich SNe & 0.0036 \\
        9  & NGC~1566    & Yes & Yes & ASASSN-14ha & 04$^{\rm{h}}$20$^{\rm{m}}$01$\fs$410  & -54\degr56\arcmin17\farcs00 & II & H-rich SNe & 0.005017 \\
        10 & NGC~1672    & Yes & Yes & SN~2017gax   & 04$^{\rm{h}}$45$^{\rm{m}}$49$\fs$430  & -59\degr14\arcmin42\farcs56 & Ib/c & SESNe & 0.00444 \\
        11 & NGC~1672    & Yes & Yes & SN~2022aau   & 04$^{\rm{h}}$45$^{\rm{m}}$41$\fs$830  & -59\degr14\arcmin43\farcs87 & II & H-rich SNe & 0.00444 \\
        12 & NGC~1729    & No  & No  & SN~2012ap    & 05$^{\rm{h}}$00$^{\rm{m}}$13$\fs$720  & -03\degr20\arcmin51\farcs20 & Ic & SESNe & 0.012155 \\
        13 & NGC~2997    & Yes & No  & SN~2003jg    & 09$^{\rm{h}}$45$^{\rm{m}}$37$\fs$910  & -31\degr11\arcmin21\farcs00 & Ib/c & SESNe & 0.003626 \\
        14 & NGC~3278    & No  & No  & SN~2009bb    & 10$^{\rm{h}}$31$^{\rm{m}}$33$\fs$870  & -39\degr57\arcmin30\farcs00 & Ic-BL & SESNe & 0.009995 \\
        15 & NGC~3354    & No  & No  & SN~2011jl    & 10$^{\rm{h}}$43$^{\rm{m}}$02$\fs$950  & -36\degr21\arcmin52\farcs40 & Ic & SESNe & 0.010001 \\
        16 & NGC~3627    & Yes & Yes & SN~1973R     & 11$^{\rm{h}}$20$^{\rm{m}}$11$\fs$500  & +12\degr59\arcmin55\farcs00 & II & H-rich SNe & 0.00193 \\
        17 & NGC~3627    & Yes & Yes & SN~1997bs    & 11$^{\rm{h}}$20$^{\rm{m}}$14$\fs$160  & +12\degr58\arcmin19\farcs56 & IIn & H-rich SNe & 0.00193 \\
        18 & NGC~3627    & Yes & Yes & SN~2009hd    & 11$^{\rm{h}}$20$^{\rm{m}}$16$\fs$990  & +12\degr58\arcmin46\farcs30 & II & H-rich SNe & 0.0024 \\
        19 & NGC~3627    & Yes & Yes & SN~2016cok   & 11$^{\rm{h}}$20$^{\rm{m}}$19$\fs$090  & +12\degr58\arcmin57\farcs20 & IIP & H-rich SNe & 0.00243 \\
        20 & NGC~3810    & No  & No  & SN~1997dq    & 11$^{\rm{h}}$40$^{\rm{m}}$55$\fs$900  & +11\degr28\arcmin45\farcs70 & Ib/c & SESNe & 0.003315 \\
        21 & NGC~4038    & No  & No  & SN~2004gt    & 12$^{\rm{h}}$01$^{\rm{m}}$50$\fs$370  & -18\degr52\arcmin12\farcs70 & Ib/c & SESNe & 0.00548 \\
        22 & NGC~4038    & No  & No  & SN~2013dk    & 12$^{\rm{h}}$01$^{\rm{m}}$52$\fs$720  & -18\degr52\arcmin18\farcs30 & Ic & SESNe & 0.00548 \\
        23 & NGC~4219    & No  & No  & SN~2011hp    & 12$^{\rm{h}}$16$^{\rm{m}}$25$\fs$470  & -43\degr19\arcmin46\farcs90 & Ic & SESNe & 0.00662 \\
        24 & NGC~4254    & Yes & Yes & SN~1967H     & 12$^{\rm{h}}$18$^{\rm{m}}$55$\fs$000  & +14\degr24\arcmin00\farcs70 & II & H-rich SNe & 0.008 \\
        25 & NGC~4254    & Yes & Yes & SN~1972Q     & 12$^{\rm{h}}$18$^{\rm{m}}$50$\fs$000  & +14\degr26\arcmin00\farcs70 & II & H-rich SNe & 0.008 \\
        26 & NGC~4254    & Yes & Yes & SN~1986I     & 12$^{\rm{h}}$18$^{\rm{m}}$52$\fs$030  & +14\degr24\arcmin43\farcs80 & II & H-rich SNe & 0.008029 \\
        27 & NGC~4254    & Yes & Yes & SN~2014L     & 12$^{\rm{h}}$18$^{\rm{m}}$48$\fs$680  & +14\degr24\arcmin43\farcs50 & Ic & SESNe & 0.008029 \\
        28 & NGC~4303    & Yes & Yes & SN~1926A     & 12$^{\rm{h}}$21$^{\rm{m}}$54$\fs$000  & +04\degr29\arcmin00\farcs60 & II & H-rich SNe & 0.00804 \\
        29 & NGC~4303    & Yes & Yes & SN~1961I     & 12$^{\rm{h}}$22$^{\rm{m}}$00$\fs$440  & +04\degr28\arcmin13\farcs30 & II & H-rich SNe & 0.0052 \\
        30 & NGC~4303    & Yes & Yes & SN~1964F     & 12$^{\rm{h}}$21$^{\rm{m}}$52$\fs$000  & +04\degr28\arcmin00\farcs40 & II & H-rich SNe & 0.0052 \\
        31 & NGC~4303    & Yes & Yes & SN~1999gn    & 12$^{\rm{h}}$21$^{\rm{m}}$57$\fs$040  & +04\degr27\arcmin45\farcs70 & IIP & H-rich SNe & 0.0052 \\
        32 & NGC~4303    & Yes & Yes & SN~2006ov    & 12$^{\rm{h}}$21$^{\rm{m}}$55$\fs$300  & +04\degr29\arcmin16\farcs70 & II & H-rich SNe & 0.0052 \\
        33 & NGC~4303    & Yes & Yes & SN~2020jfo   & 12$^{\rm{h}}$21$^{\rm{m}}$50$\fs$480  & +04\degr28\arcmin54\farcs05 & IIP & H-rich SNe & 0.00522 \\
        34 & NGC~4321    & Yes & Yes & SN~1979C     & 12$^{\rm{h}}$22$^{\rm{m}}$58$\fs$580  & +15\degr47\arcmin52\farcs70 & II & H-rich SNe & 0.00455 \\
        35 & NGC~4321    & Yes & Yes & SN~2020oi    & 12$^{\rm{h}}$22$^{\rm{m}}$54$\fs$925  & +15\degr49\arcmin25\farcs05 & Ic & SESNe & 0.00524 \\
        36 & NGC~4981    & No  & No  & SN~2007C     & 13$^{\rm{h}}$08$^{\rm{m}}$49$\fs$300  & -06\degr47\arcmin01\farcs00 & Ib & SESNe & 0.005595 \\
        37 & NGC~5128    & No  & No  & SN~2016adj   & 13$^{\rm{h}}$25$^{\rm{m}}$24$\fs$110  & -43\degr00\arcmin57\farcs90 & IIb & SESNe & 0.001889 \\
        38 & NGC~6744    & Yes & No  & SN~2005at    & 19$^{\rm{h}}$09$^{\rm{m}}$53$\fs$570  & -63\degr49\arcmin22\farcs80 & Ic & SESNe & 0.002803 \\
        39 & PGC~037625  & No  & No  & SN~2014ad    & 11$^{\rm{h}}$57$^{\rm{m}}$44$\fs$440  & -10\degr10\arcmin15\farcs70 & Ic-pec & SESNe & 0.0053 \\
        40 & PGC~084885  & No  & No  & SN~2004ip    & 18$^{\rm{h}}$32$^{\rm{m}}$41$\fs$260  & -34\degr11\arcmin26\farcs70 & II & H-rich SNe & 0.018276 \\
        \hline  
    \end{tabular}
\end{table*}

The results obtained in our work are given in Table~\ref{tab:tabC2}.
Specifically, at the positions of CCSNe we measured the star formation rate surface density, the molecular gas surface density, and the star formation efficiency.

\begin{table*}
    \caption{CCSN environments characterised in this work.}
    \label{tab:tabC2}
    \centering
    \begin{tabular}{r|cccccc}
        \hline
        {} & SN name & SN group & $\Sigma_{\mathrm{SFR}}$ & $\Sigma_{\mathrm{mol}}$ & SFE & ALMA detection\\
        {} & {}      & {}       & $\left[ \rm{M}_{\odot}\,yr^{-1}\,kpc^{-2}\right]$ & $\left[ \rm{M}_{\odot}\,kpc^{-2} \right] $ & $\left[\rm{Gyr}^{-1}\right]$ & {}\\
        {} & {}      & {}       & $\times 10^{-5}$ & $\times 10^{6}$ & $\times 10^{-2}$ & {}\\
        \hline
        1 & ASASSN-14ha & H-rich SNe & $1698 \pm 17$     & $871 \pm 28$ & $2 \pm 0.1$ & Detection \\
        2 & LSQ13doo    & SESNe      & $1431 \pm 35$     & $10 \pm 6$   & $142 \pm 8$ & Non-detection \\
        3 & SN~1926A     & H-rich SNe & $467 \pm 44$      & $27 \pm 14$  & $18 \pm 9$ & Non-detection \\
        4 & SN~1961I     & H-rich SNe & $1102 \pm 32$     & $55 \pm 28$  & $20 \pm 10$ & Non-detection \\
        5 & SN~1964F     & H-rich SNe & $171 \pm 11$      & $39 \pm 21$  & $4 \pm 2$ & Non-detection \\
        6 & SN~1967H     & H-rich SNe & $5531 \pm 42$     & $215 \pm 8$  & $26 \pm 1$ & Detection \\
        7 & SN~1972Q     & H-rich SNe & $498 \pm 11$      & $54 \pm 9$   & $9 \pm 2$ & Detection \\
        8 & SN~1973R     & H-rich SNe & $126 \pm 5$       & $155 \pm 16$ & $1 \pm 0.1$ & Detection \\
        9 & SN~1979C     & H-rich SNe & $222 \pm 59$      & $23 \pm 12$  & $10 \pm 6$ & Non-detection \\
        10 & SN~1983V    & SESNe      & $1885 \pm 12$     & $56 \pm 7$   & $34 \pm 4$ & Detection \\
        11 & SN~1985P    & H-rich SNe & $56 \pm 9$        & $28 \pm 18$  & $2 \pm 1$ & Non-detection \\
        12 & SN~1986I    & H-rich SNe & $2295 \pm 32$     & $98 \pm 12$  & $23 \pm 3$ & Detection \\
        13 & SN~1994ai   & SESNe      & $4860 \pm 110$    & $2 \pm 1$    & $2400 \pm 1200$ & Non-detection \\
        14 & SN~1995V    & H-rich SNe & $1023 \pm 10$     & $179 \pm 13$ & $6 \pm 0.4$ & Detection \\
        15 & SN~1997bs   & H-rich SNe & $50 \pm 6$        & $24 \pm 16$  & $2 \pm 2$ & Non-detection \\
        16 & SN~1997dq   & SESNe      & $806 \pm 16$      & $56 \pm 32$  & $14 \pm 8$ & Non-detection \\
        17 & SN~1999gn   & H-rich SNe & $5817 \pm 33$     & $132 \pm 11$ & $44 \pm 4$ & Detection \\
        18 & SN~2001du   & H-rich SNe & $730 \pm 9$       & $35 \pm 8$   & $21 \pm 5$ & Detection \\
        19 & SN~2003jg   & SESNe      & $6290 \pm 130$    & $49 \pm 14$  & $128 \pm 37$ & Detection \\
        20 & SN~2004gt   & SESNe      & $12410 \pm 220$   & $76 \pm 38$  & $163 \pm 82$ & Non-detection \\
        21 & SN~2004ip   & H-rich SNe & $508600 \pm 7500$ & $127 \pm 35$ & $4000 \pm 1100$ & Detection \\
        22 & SN~2005at   & SESNe      & $1901 \pm 38$     & $675 \pm 15$ & $3 \pm 0.1$ & Detection \\
        23 & SN~2005lr   & SESNe      & $19910 \pm 280$   & $53 \pm 5$   & $375 \pm 36$ & Detection \\
        24 & SN~2006ov   & H-rich SNe & $1348 \pm 86$     & $33 \pm 17$  & $41 \pm 21$ & Non-detection \\
        25 & SN~2007C    & SESNe      & $5480 \pm 160$    & $47 \pm 3$   & $117 \pm 8$ & Detection \\
        26 & SN~2009bb   & SESNe      & $77200 \pm 1100$  & $110 \pm 30$ & $700 \pm 190$ & Detection \\
        27 & SN~2009hd   & H-rich SNe & $2143 \pm 12$     & $479 \pm 15$ & $4 \pm 0.2$ & Detection \\
        28 & SN~2011hp   & SESNe      & $11780 \pm 250$   & $21 \pm 6$   & $550 \pm 160$ & Detection \\
        29 & SN~2011jl   & SESNe      & $1134 \pm 30$     & $6 \pm 3$    & $186 \pm 93$ & Non-detection \\
        30 & SN~2012ap   & SESNe      & $1201 \pm 72$     & $15 \pm 9$   & $80 \pm 46$ & Non-detection \\
        31 & SN~2013dk   & SESNe      & $13010 \pm 240$   & $32 \pm 16$  & $410 \pm 210$ & Non-detection \\
        32 & SN~2014L    & SESNe      & $7547 \pm 49$     & $393 \pm 15$ & $19 \pm 1$ & Detection \\
        33 & SN~2014ad   & SESNe      & $4673 \pm 66$     & $4 \pm 2$    & $1090 \pm 540$ & Non-detection \\
        34 & SN~2016adj  & SESNe      & $20970 \pm 530$   & $69 \pm 4$   & $305 \pm 21$ & Detection \\
        35 & SN~2016cok  & H-rich SNe & $400 \pm 5$       & $81 \pm 10$  & $5 \pm 1$ & Detection \\
        36 & SN~2017gax  & SESNe      & $513 \pm 26$      & $114 \pm 15$ & $5 \pm 1$ & Detection \\
        37 & SN~2020jfo  & H-rich SNe & $816 \pm 12$      & $29 \pm 15$  & $28 \pm 14$ & Non-detection \\
        38 & SN~2020oi   & SESNe      & $10308 \pm 57$    & $216 \pm 18$ & $48 \pm 4$ & Detection \\
        39 & SN~2022aau  & H-rich SNe & $10243 \pm 31$    & $497 \pm 20$ & $21 \pm 1$ & Detection \\
        40 & SN~2022acko & H-rich SNe & $204 \pm 36$      & $11 \pm 6$   & $19 \pm 11$ & Non-detection \\
        \hline
\end{tabular}
\end{table*}

%%%%%%%%%%%%%%%%%%%%%%%%%%%%%%%%%%%%%%%%%%%%%%%%%%

% Don't change these lines
\bsp	% typesetting comment
\label{lastpage}
\end{document}